%% file: 1.main-IEEE.tex
\documentclass[conference]{IEEEtran}

\usepackage[utf8]{inputenc}
\usepackage[T1]{fontenc}
\usepackage{amsmath,amssymb,amsfonts}
\usepackage{graphicx}
\usepackage{textcomp}
\usepackage{booktabs}
\usepackage{flafter}
\usepackage{tikz}
\usepackage{listings}
\usepackage{pifont}
\usepackage{multirow}
\usepackage{colortbl}
\usepackage{soul}
\usepackage{paralist}
\usepackage{enumitem}
\usepackage[sort&compress,numbers]{natbib}

\usepackage[most]{tcolorbox}
\tcbset{
    colback=white,
    colframe=black,
    boxrule=0.8pt,
    arc=6pt,
    left=2pt,
    right=2pt,
    top=2pt,
    bottom=2pt
}

\usepackage{xcolor}
\usepackage{hyperref}
\hypersetup{
    colorlinks=true,
    unicode=true,
    linkcolor=[rgb]{0.10,0.05,0.67},
    citecolor=[rgb]{0.10,0.05,0.67},
    filecolor=[rgb]{0.10,0.05,0.67},
    urlcolor=[rgb]{0.10,0.05,0.67}
}

\makeatletter
\let\orig@Hy@StartLink\Hy@StartLink
\def\Hy@StartLink#1#2{%
    \ifx#1\relax
    \else
        \orig@Hy@StartLink{#1}{#2}%
    \fi
}
\makeatother

\newcommand{\mytitle}{EIR}

\title{\underline{E}mulated \underline{I}ntegrity \underline{R}eplica:
Enabling Self-Healing on FPGA SoCs via Hierarchical Twins}

\author{
\IEEEauthorblockN{
Arsalan Ali Malik\IEEEauthorrefmark{1},
Ali Suvizi\IEEEauthorrefmark{2},
Guru Venkataramani\IEEEauthorrefmark{2},
Aydin Aysu\IEEEauthorrefmark{1}
}
\IEEEauthorblockA{
\IEEEauthorrefmark{1}North Carolina State University,
Raleigh, NC, USA\\
\IEEEauthorrefmark{2}George Washington University,
Washington, DC, USA\\
{\{aamalik3, aaysu\}@ncsu.edu},
{\{ali.suvizi, guruv\}@gwu.edu}
}
}

\begin{document}

\maketitle

\thispagestyle{plain}
\pagestyle{plain}

% ...
\input{0.abstract}


\input{2.introduction}
\input{3.background_preliminary}
\input{4.methodology}
\input{5.results}
\input{6.discussions}
\input{7.conclusion}

{\fontsize{7.5pt}{8.5pt}\selectfont
\bibliographystyle{IEEEtran}
\bibliography{References}
}

\end{document}

%% file: 0.abstract.tex
\begin{abstract}
Convolutional neural networks (CNNs) are increasingly being deployed on system-on-chip (SoC) platforms, where hardware-accele\-rated inference enables low-latency edge computing. Achieving fault tolerance on these devices remains challenging because conventional redundancy (dual/triple modular redundancy, DMR/TMR) incurs high resource cost, while software-centric methods—\textit{e.g.,} algorithm-based fault tolerance (ABFT), checkpoint–restart, instruc\-tion-level duplication, and software watchdogs/assertions—introdu\-ce nontrivial latency/energy overheads, reduce model accuracy, or provide inadequate coverage for accelerator-induced faults. 

In this paper, we propose \textit{Emulated Integrity Replicate (EIR)}, a hierarchical digital-twin framework for FPGA SoCs that provides autonomous fault detection and recovery. Unlike DMR/TMR, which replicates hardware logic and incurs proportional area and power overheads, ~\mytitle~ avoids fabric-level duplication by exploiting temporal slack in the processing system (PS). During accelerator execution in the programmable logic (PL), the PS typically remains underutilized; ~\mytitle~ capitalizes on these idle cycles to host two complementary twins: (i) \texttt{Rabbit}—a coarse-grained behavioral model for rapid fault detection and (ii) \texttt{Tortoise}—a fine-grained gate-level model that performs precise recovery from checkpointed states. The accelerator state is captured periodically, leveraging the accelerator's execution-speed profiling to balance performance overhead and resilience. Experiments on representative workloads show that~\mytitle~ achieves high empirical fault coverage relative to a DMR baseline while reducing energy and area under the evaluated fault model and workload assumptions, indicating a practical path to resilient edge-AI deployments under strict resource budgets.

\end{abstract}


%% file: 2.introduction.tex
\begin{figure}[t]
    \vspace{0.35em}
	\includegraphics[width=0.48\textwidth]{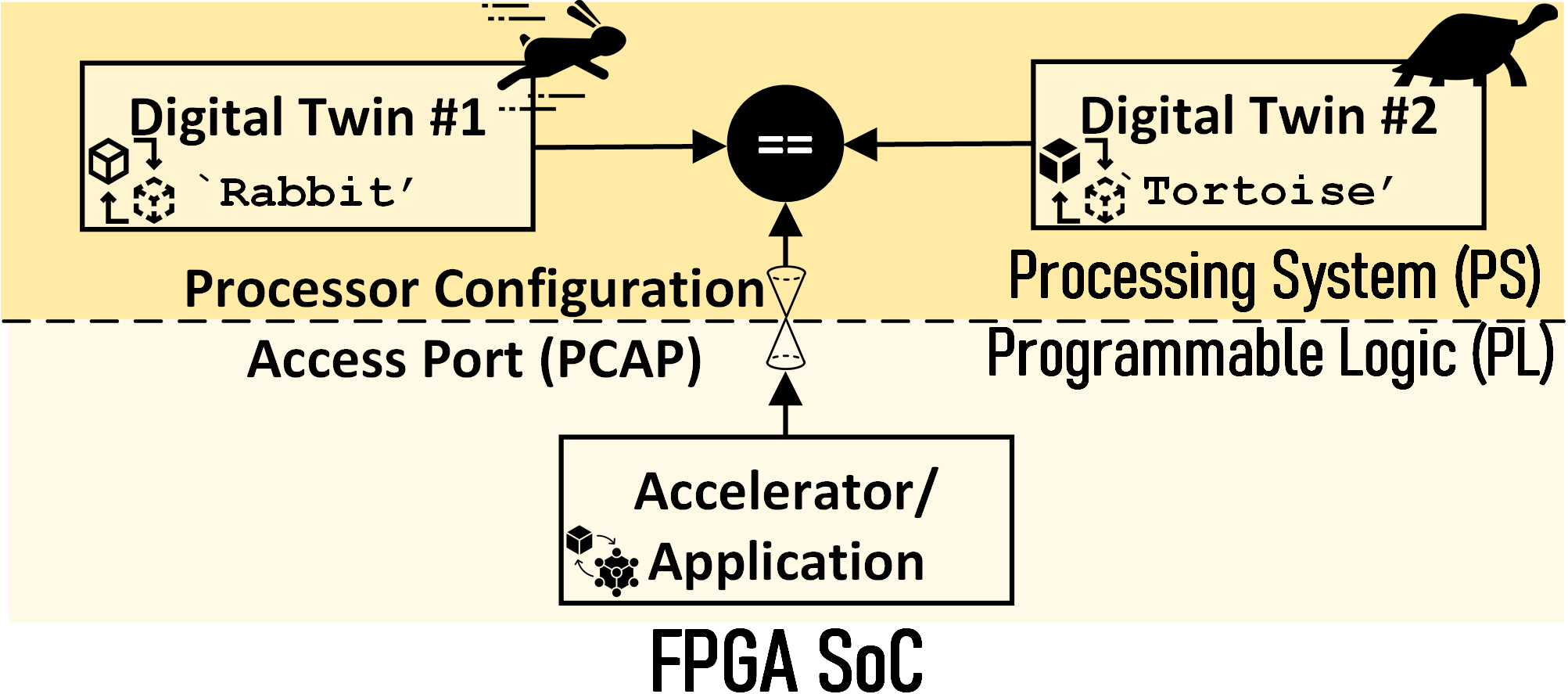}
	\caption{\mytitle~ is a self-healing framework for resource-constrained FPGA SoCs at the edge that leverages idle cycles in the PS to host two coordinated digital twins for an accelerator in the PL. Digital Twin~\#1, \texttt{Rabbit}, is a fast, coarse-grained behavioral model that runs continuously to detect divergence from the PL application. Digital Twin~\#2, \texttt{Tortoise}, is a slow, fine-grained gate-level–equivalent model that provides precise recovery when needed. When \texttt{Rabbit} detects a mismatch, the latest fault-free PL state is loaded into \texttt{Tortoise}, which deterministically advances the design to a known-good state; this state is then written back to the PL to recover and resume correct execution without DMR/TMR overheads in hardware.}
\vspace{-1.0em}
	\label{fig: Concept}
\end{figure}
\section{Introduction}
Edge AI deployments in safety-critical domains (\textit{e.g., }autonomous vehicles, industrial automation, etc.) require fault tolerance under tight hardware and environmental constraints~\cite{antunes2025fpga,tesla2016autopilot,suvizi2025autohealer,malik2025precision,wirthlin2015high}. Dual and triple modular redundancy (DMR/TMR) remain the de facto standard in such systems~\cite{DMR,TMR,taheri2024adam,ghionda2025hmr}, \textit{e.g.,} Tesla’s full self-driving~\cite{tesla2016autopilot} and OpenTitan’s root of trust~\cite{opentitan2019}. DMR (at least) doubles, and TMR triples, the hardware footprint, increasing area and power and constraining resources for the primary application.

Software-based fault tolerance offers a lower-cost alternative to hardware redundancy. Examples include checkpoint-and-rollback, which periodically saves system state and restores it upon fault detection~\cite{teodorescu2006cache}, and algorithm-based fault tolerance (ABFT)
, which embeds checksums into neural network (NN) computations to detect errors without full hardware replication ~\cite{ABFT_zhao2020ft,software_checkpointteodorescu2005empowering}. On FPGA SoCs, however, checkpointing requires dedicated support for efficient state capture that commercial devices do not provide~\cite{teodorescu2006cache,attia2020statemover}, and ABFT has not yet been integrated with FPGA accelerators to achieve complete fault coverage~\cite{xue2023approxabft}. Moreover, both approaches lack the fine-grained error localization needed for precise recovery.

Fig.~\ref{fig: Concept} outlines our proposed framework~\mytitle~, which combines digital twin notions with FPGA SoC hardware-software co-design to enable self-healing at a system level.
The key novelty of~\mytitle~is its hierarchical digital-twin architecture that (i) exploits idle processing system (PS) cycles, (ii) avoids DMR/TMR area overhead, (iii) uses a checkpoint policy, and (iv) incorporates a state-capture-and-restoration mechanism to opportunistically snapshot programmable logic (PL) state into off-chip memory.
This digital twin hierarchy consists of: (i) \texttt{Rabbit}, a fast (coarse-grained) behavioral model that monitors execution and detects deviations (of its PL counterpart) by recomputing at a higher-level abstraction in software. This twin prioritizes
speed over cycle-accurate fidelity, enabling rapid periodic
validation of accelerator outputs to detect faults. (ii) \texttt{Tortoise}
, a slower (fine-grained)  gate-level equivalent model that remains idle until a fault is detected. When activated, \texttt{Tortoise} emulates execution from a \textit{known-good} checkpoint and reconstructs a correct system state for on-chip recovery.

To address the inherent speed and resource imbalance between the PS and the accelerator, ~\mytitle~employs a latency-aware checkpointing policy that twins only critical computation phases. At selected checkpoints, \texttt{Rabbit} executes the same inputs as the PL accelerator and compares outputs. Upon detecting a mismatch, \texttt{Rabbit} triggers \texttt{Tortoise} to replay a fault-free checkpoint at gate-level fidelity to emulate the correct system state\footnote{\texttt{Tortoise} always ``wins'' on faults but \texttt{Rabbit} enables faster detection; unlike in tales, here they work together to drive system toward fault-free operation.}. PL then restores this state, enabling software-like checkpoint-based recovery without DMR/TMR hardware redundancy.
~\mytitle’s hierarchical approach offers three advantages over conventional redundancy: (i) it activates only during detection and recovery instead of running redundant hardware continuously, (ii) it avoids duplicating PL accelerator logic by hosting twins on the PS and storing checkpoints in off-chip memory, and (iii) it supports mid-execution state restoration through the gate-level twin, whereas DMR typically halts operation without in-situ recovery. 

The key contributions of this work are:



\begin{itemize}[noitemsep, nolistsep]
    \item Hierarchical digital twin architecture: We introduce the \textit{first} dual-granularity digital twin framework specifically designed for FPGA SoC accelerators, combining rapid functional detection with precise gate-level recovery.

    \item Hardware-software checkpoint integration: We demonstrate a practical implementation of checkpoint-based fault tolerance leveraging FPGA SoC-specific features with execution speed-informed checkpoint selection that balances detection latency against storage overhead.

    \item Self-healing methodology for edge AI: We present an end-to-end system enabling autonomous fault recovery in resource-constrained accelerators, offering an alternative to full redundancy that delivers comparable empirical fault coverage at reduced energy and area cost.
    

    \item Experimental validation: We demonstrate~\mytitle's applicability on representative neural network topologies, achieving top $30$\% neuron coverage with $\geq95$\% fault detection rate, while quantifying its resource benefits against DMR baselines. Moreover, we evaluated $11$ applications as \emph{representative} benchmarks. Across the tested accelerators,~\mytitle~preserves the baseline PL area, whereas DMR increases LUT/FF utilization by approximately $2.0$--$2.25\times$ and raises power consumption from $2.98$~W to $5.79$~W ($\approx89.1$\%), on average.
\end{itemize}

Unlike prior works that either replicate hardware or detect errors without recovery,~\mytitle~ is the \textit{first} to combine software-based detection with cycle-accurate state reconstruction for in-place recovery on FPGA SoCs.~\mytitle~'s hierarchical twin design allows fault tolerance to be achieved without modifying accelerator RTL or duplicating logic in the reconfiguration fabric.



%% file: 3.background_preliminary.tex
\section{Background and Related Work}
This section reviews prior efforts in fault tolerance, redundancy reduction, and FPGA SoC resilience, most relevant to~\mytitle.
\subsection{Hardware Redundancy for Fault Tolerance}
Spatial redundancy is a common technique for fault tolerance~\cite{antunes2025fpga,taheri2024adam}. TMR replicates a module three times with majority voting to mask single faults. TMR achieves high coverage but consumes $2$-$3\times$ more resources and increases power~\cite{TMR}. DMR reduces this overhead using two replicas, but lacks automatic correction~\cite{DMR}. Recent work proposed HMR-NEureka, a selective scheme for FPGA neural accelerators~\cite{ghionda2025hmr}. However, it still incurs sizable overheads—reporting a $2.1\times$ area increase for a RISC-V core with integrated error prediction. Hierarchical designs, such as HyCA reduce cost by pairing $2$D processing element (PE) array with a backup unit that recomputes operations mapped to faulty PEs, but they still require dedicated spare hardware in the fabric~\cite{HyCA}. In edge AI deployments with stringent area and power budgets, even partial hardware redundancy, such as selective replication of critical components, can be impractical due to its resource overhead. By contrast,~\mytitle~sidesteps fabric-level duplication \emph{entirely} by repurposing underutilized PS to achieve comparable fault resilience via software-based replication.


\subsection{Algorithm-Based Fault Tolerance (ABFT)}

ABFT uses mathematical structure of computations to embed fault detection and correction capabilities without full hardware duplication~\cite{taheri2024adam}. For NNs, ABFT typically augments matrix operations with row and column checksums computed during forward propagation in which mismatches between output and expected checksums indicate errors. ApproxABFT extends this concept by introducing approximate, tolerance-aware checksums and reports a $43.39$\% reduction in overhead compared to traditional ABFT while increasing tolerable error rates~\cite{xue2023approxabft}.
However, deploying ABFT on FPGA SoCs faces three challenges: (i) architecture-agnostic designs miss hardware-specific efficiency opportunities, (ii) checksum computation and verification may violate edge inference latency budgets, and (iii) ABFT primarily detects
, but does not correct, faults~\cite{ABFT_zhao2020ft}. For mission-critical applications, this detection-only capability is insufficient, as it neither prevents silent data corruption nor restores correct operation; instead, timely fault localization and automated state recovery are required to maintain correct operation.



\subsection{Checkpoint and Rollback Mechanisms}
Checkpoint-based fault tolerance periodically saves system state to stable storage and restores it upon errors, enabling backward recovery without full re-execution~\cite{software_checkpointteodorescu2005empowering,teodorescu2006cache,malik2025epoch}. While effective in processor-centric systems, applying checkpointing to FPGA SoC accelerators is non-trivial because: (i) Accelerator state spans registers, distributed memories, pipeline buffers, and configuration bits, and (ii) checkpoint intervals must balance detection latency (longer intervals increase exposure to undetected faults) against memory traffic and storage overhead.
Recent FPGA debugging frameworks (\textit{e.g.,} StateMover~\cite{attia2022stop}, Zoomie~\cite{wei2024zoomie}) combine FPGA execution speed with simulation observability using checkpoint-based flows~\cite{attia2020statemover,attia2021statemover_github}. These approaches, however, assume PC-based host-in-the-loop simulation and target pre-deployment verification. By contrast,~\mytitle: (i) pursues autonomous, in-field fault tolerance rather than verification workflows, (ii) uses hierarchical twins—for runtime fault-detection and recovery,  and (iii) runs entirely on the chip, enabling self-healing at the edge, avoiding host dependence.

\subsection{Existing Self-Healing Architectures}
Self-healing FPGA systems detect and repair configuration upsets by combining scrubbing—periodically reading and correcting configuration memory—with partial reconfiguration (PR) to replace faulty regions~\cite{stoddard2016high,nepalharnessing}. Scrubbing fixes only configuration-memory faults, not logic errors in on-chip data or registers. Early PR-based systems recover from single-event upsets (SEUs) but depend on an external host (often a PC) to regenerate bitstreams, reducing autonomy and increasing latency and complexity~\cite{custodio2007selfhealingPR}. Scalable self-healing circuits automate recovery but focus on control logic, leaving silent data corruptions and subtle computation errors in accelerator datapaths unresolved~\cite{Scalable_and_Accelerated_Selfhealing}. Auto-Healer detects perception faults in autonomous driving by monitoring runtime behavior and triggering recovery~\cite{suvizi2025autohealer}. Although aligned with~\mytitle's goals, Auto-Healer operates at the software algorithm level (\textit{e.g.}, misclassification detection) and relies on DMR-style replication for error detection. By contrast, ~\mytitle~removes fabric-level redundancy and uses hierarchical digital twins to detect and repair circuit-level faults directly, preventing them from propagating into misclassifications.

\begin{figure}[t!]
	    
     \includegraphics[width=0.48\textwidth]{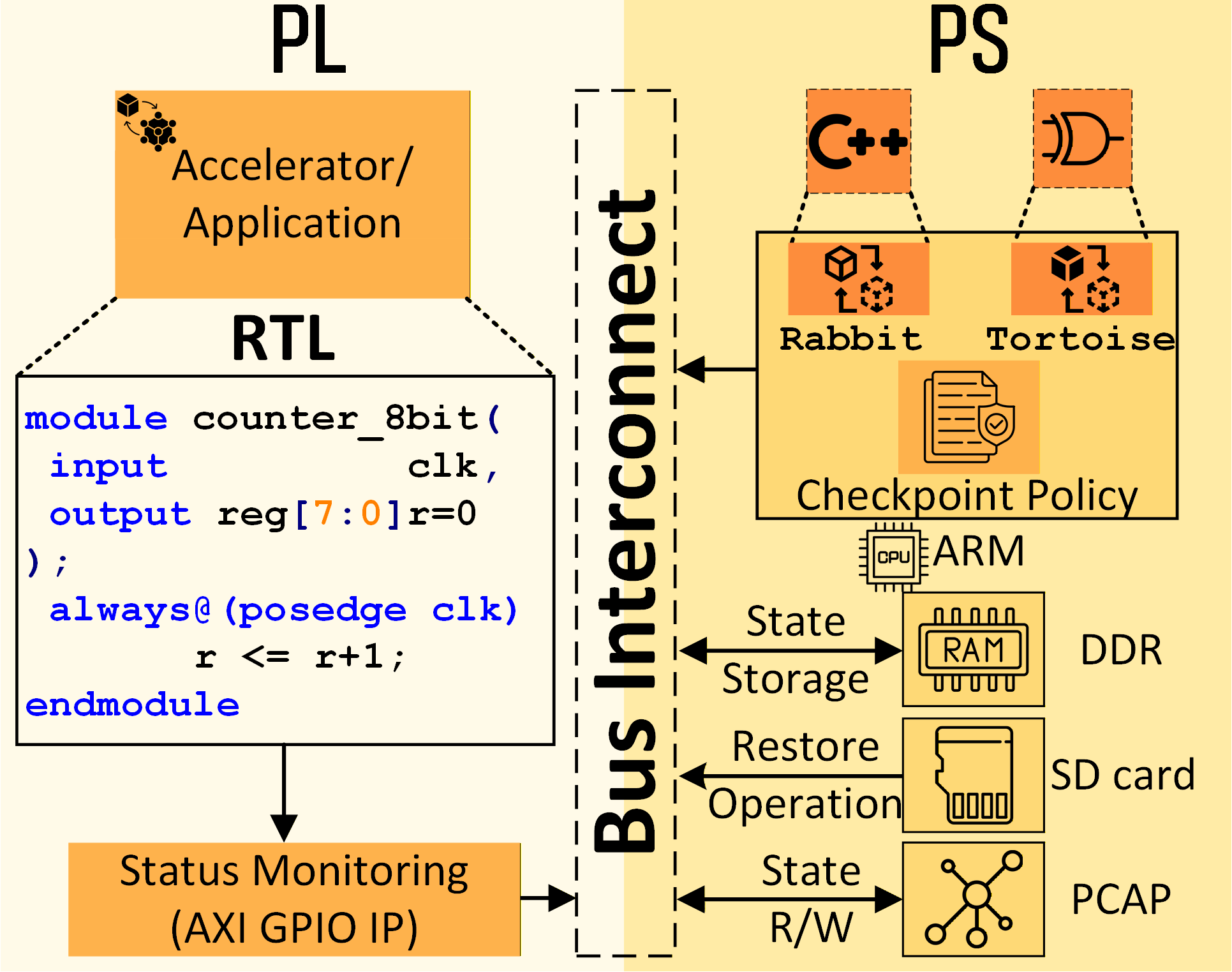}
    \caption{\mytitle’s system-level architecture for an FPGA SoC runs on the PS, while the accelerator resides in the PL. An AXI-GPIO IP monitors accelerator outputs and flags faults; detection and recovery in the PS avoid fabric-level replicas and preserve PL resources. PL state is captured via PCAP with no PL area overhead and selectively read back into off-chip DDR using logic-location (*.ll) mappings stored on an SD card. The PS hosts the checkpoint policy and two digital twins: \texttt{Rabbit}, a C/C++ behavioral twin for output comparison, and \texttt{Tortoise}, a CXXRTL-generated C++ gate-level twin for cycle-accurate localization and checkpoint-based recovery.
    \vspace{1.5em}
    }

\vspace{-2.75em}
	\label{fig: System}
\end{figure}
\subsection{~\mytitle's Scope and Assumptions}
\vspace{-.25em}
~\mytitle~targets FPGA SoC platforms, \textit{e.g.,} AMD/Xilinx Zynq-$7000$/Ultra\-Scale/UltraScale+ devices. The primary accelerator executes in the PL, while the PS runs the application's twins and hosts the~\mytitle~framework.~\mytitle~can mitigate both single and multi-bit transient faults caused by voltage fluctuations, electromagnetic interference, and radiation-induced bit flips that may affect (i) PL configuration memory, (ii) on-chip data memories (storing weights and biases\footnote{We exclude on-chip block RAM (BRAM) from scope because AMD/Xilinx devices provide built-in error-correcting code (ECC) for BRAM, offering meaningful protection against soft errors. This allows us to focus on faults in logic and interconnect, where no comparable native protection exists.}). We assume Block-RAMs, PS, and off-chip checkpoint memory (DDR) are protected by error-correcting code (ECC) or similar mechanisms and thus remain trustworthy for hosting the twins and coordinating recovery~\cite{salami2020ecc}. Permanent faults are outside~\mytitle's scope, focusing on transient errors that can be corrected by restoring a \textit{known-good} state.

\mytitle~is best suited to digital accelerators whose relevant state can be checkpointed and replayed without irreversible external side effects, and platforms where PS can run \texttt{Rabbit} at checkpoints and \texttt{Tortoise} on rare faults.~\mytitle~is well-suited to applications that demand lower PL area and energy overhead and can accommodate bounded recovery stalls\textit{ e.g.,} edge AI and cloud FPGAs ~\cite{karabulut2024themis}.

%% file: 4.methodology.tex
\section{The Proposed Framework:~\mytitle}\label{Sec: HW_design}
Fig.~\ref{fig: System} shows~\mytitle’s system-level architecture. The PL hosts the target accelerator, while an AXI-GPIO core monitors its outputs. Fault detection and recovery are assigned to the PS, preserving precious PL resources by avoiding fabric-level replicas.~\mytitle~uses the PCAP to read/write PL state, leveraging a logic-location (*.ll) map stored on an attached SD card to selectively read back \emph{only} utilized regions into off-chip DDR memory. The PS hosts the two digital twins and the checkpoint policy described in Sections~\ref{Subsec: Hierarchical Twins: Architecture and Flow} and~\ref{subsec: Checkpoint Selection Policy}, respectively.


\subsection{Hierarchical Twins: Architecture and Flow}
\label{Subsec: Hierarchical Twins: Architecture and Flow}
\texttt{Rabbit} (Twin \#1) is a coarse-grained twin that operates at a higher abstraction level, replicating the functional behavior of the PL accelerator without modeling individual clock cycles or hardware signal transitions. It executes on the PS, producing expected outputs based on equivalent computational logic, written in C/C++. This higher level abstraction enables fast execution, rapid fault detection and consistency verification—\texttt{Rabbit} (typically) completes its operation faster than a cycle-accurate simulation would require~\cite{attia2020statemover,wei2024zoomie}. However, \texttt{Rabbit} cannot shadow the PL accelerator continuously, as in DMR, for three reasons: (i) Typically, the PS executes an application slower than the PL, so a full, always-on replica would lag behind, (ii) continuously running \texttt{Rabbit} would monopolize PS cycles needed for other system tasks, and (iii) persistent execution would increase energy consumption, conflicting with edge constraints. To address this inherent speed and resource imbalance,~\mytitle~employs a latency-aware decision mechanism that selectively identifies critical computation phases to twin (refer Sec.~\ref{subsec: Checkpoint Selection Policy}). At these checkpoints, \texttt{Rabbit} processes the same input batch as the PL accelerator and compares outputs. Upon a mismatch, it triggers \texttt{Tortoise} for fault mitigation and system recovery.


\texttt{Tortoise} (Twin \#2) is a fine-grained, gate-level twin that provides cycle-accurate emulation of the PL accelerator. Tortoise uses the Yosys CXXRTL backend to convert the same RTL used for the PL bitstream into a C++ model that runs on the PS~\cite{YosysCXXRTL}. This preserves registers, combinational logic, and signal level details, allowing \texttt{Tortoise} to reproduce the hardware’s internal state evolution. Although slower than \texttt{Rabbit} or native PL execution, \texttt{Tortoise's} fidelity makes it suitable for precise fault localization and recovery. Starting from a verified checkpoint, it can emulate forward to compute the correct PL state and regenerate fault-free outputs.~\mytitle~assumes \texttt{Rabbit} consumes only a fraction of available PS cycles during steady-state execution, and that \texttt{Tortoise} is invoked infrequently through exception mechanisms only when needed. The effectiveness of \texttt{Rabbit} depends on the availability of PS compute resources. As PS utilization increases—due to application or system-level workloads—the fraction of computation that can be validated online \textit{may} decrease. While this \textit{may} increase fault-detection latency, it does \emph{not} affect the correctness of recovery once a fault is detected. In such cases, coarser checkpointing or more selective validation policies can be employed to maintain system responsiveness.
\subsection{Checkpoint Selection Policy}
\label{subsec: Checkpoint Selection Policy}
Running programs on PS instead of PL could slow it down considerably, making continuous cycle-accurate shadowing infeasible. Guided by prior work and our LeNet-$5$ case study,~\mytitle~instead uses a policy that targets decision-critical computations and places checkpoints around them to bound divergence. We used an AI workload for characterization because neural networks are both highly error-sensitive and representative of modern edge deployments on FPGA SoCs, especially in safety-critical domains, such as autonomous driving, robotics, and industrial automation~\cite{amd_automotive_zynq_7000_xa,xilinx_logiADAK_adas,tesla2016autopilot}. Moreover, neural networks exhibit highly \textit{non-uniform }fault sensitivity across layers, making them a realistic and rigorous benchmark for evaluating~\mytitle’s fault detection, localization, and recovery capabilities~\cite{rakin2019bit,gongye2024one}.


We derive the selection policy in two stages: (i) literature informed sensitivity analysis to identify where faults have the greatest impact~\cite{b1,b2,b3,gongye2024one,rakin2019bit}, and (ii) a model-specific case study to quantify how much comparison is sufficient. Prior work indicates that (a) faults in early convolutional layers are often masked, while deeper fully connected and output (logit) layers are likely to corrupt outputs~\cite{b1,b4}, (b) vulnerability concentrates in high-magnitude multiply-accumulate (MAC) operations, so protecting high-activation regions yields the largest resilience gain~\cite{b2,b7,b6}, and (c) in fixed-point designs, higher-significance bits (\textit{e.g.,} MSBs) dominate fault impact~\cite{b5}. While neural networks exhibit inherent error masking, our sensitivity analysis (see Sec~\ref{sec: Fault Coverage }) shows that a small subset of high-contribution MACs dominates outcome-altering faults. This observation motivates ~\mytitle’s selective validation strategy, which targets these critical regions rather than uniformly protecting all computations, enabling efficient fault coverage despite the presence of masking effects.

The checkpoint and validation policy used in this work is intentionally profiling-guided. It is designed to prioritize computation phases that are most likely to affect the final output under the evaluated workload, rather than to provide architecture-independent or formally complete coverage. For workloads with different fault-propagation behavior, dynamic control flow, or shifting input distributions, the protected subset and checkpoint boundaries may need to be re-profiled or selected adaptively.
Based on these insights, we next detail our case study.

\begingroup
\definecolor{CaseStudyColor}{HTML}{000000} 
\color{CaseStudyColor}
\subsubsection{\textbf{Model-specific case study}}
We implement the LeNet-5 architecture for image classification and pre-train it on CIFAR-10 to obtain inference weights `$w$'. After training, we apply post-training quantization to the $w$ and activations to reduce resource usage and align with the deployed accelerator’s precision. The model includes eight layers—two convolutional, two pooling, one flatten, and three fully connected (dense) layers. We describe the design in VHDL at the RTL and synthesize it for FPGA deployment.

\subsubsection{\textbf{Benign baseline and fault injection.}}
For each input `$x$', we begin with a fault-free execution and record (i) the per-layer outputs and (ii) the final predicted class to establish a reference label and a per-layer catalogue of MAC contribution magnitudes $|x\!\cdot\!w|$. We then apply a SEU fault model at the MAC-operation level: for a target layer and MAC instance `$i$', we inject bit flip
(s) into the MAC result register at the cycle of its generation. We perform runtime fault injection through testbench simulations to assess the system’s fault tolerance.
\subsubsection{\textbf{Stage~1 (layer-level sensitivity).}}\label{Stage_1}
To examine whether different layers contribute equally to output corruption, we conduct per-layer fault injection campaigns that randomly select MAC instances within the target layer and inject bit-flip(s) per inference. For each injection, we compare the resulting classification with the fault-free baseline to determine whether it causes a misclassification. Aggregating results across all injections produces a per-layer misclassification profile, revealing that convolutional layers generally mask faults, whereas fully connected (dense) and logit layers are more likely to alter the final decision.
\subsubsection{\textbf{Stage~2 (magnitude-aware sensitivity)}}\label{Stage_2}
Building on the observation that dense layers have the greatest impact on output integrity, we next quantify which MAC operations within each dense layer contribute most to fault susceptibility. For each dense layer $l$, input $x$, and corresponding weight $w$, we compute the absolute MAC contributions as follows:
\[
m_{l,i}(x) \;=\; \big|\,x_{l,i}\cdot w_{l,i}\,\big|
\]

We rank all MAC operations in descending order of $|MAC_{l,i}|$ and divide them into contiguous percentiles (\textit{e.g.}, deciles) to study how fault sensitivity varies with contribution magnitude. We then sample MAC indices from each percentile and inject single or multi-bit faults as before. For each percentile, we record misclassifications and class transitions. Aggregating results across percentiles shows that faults cluster in the top ranges, revealing a compact top-$k_l$ subset in each dense layer responsible for most misclassifications.


\endgroup

\vspace{-.5em}
\subsection{System Checkpoint and Rollback}
\label{Subsec:EPOCH}
\vspace{-.25em}
~\mytitle~leverages PCAP to read/write the PL state, building on~\cite{malik2025epoch}, enabling fine-grained state readback of heterogeneous FPGA elements with safe clock gating and synchronized resumption, which directly informs~\mytitle’s checkpointing strategy\footnote{We used the approach in~\cite{malik2025epoch} to implement checkpointing in~\mytitle~because (i) unlike StateMover~\cite{attia2020statemover} and Zoomie~\cite{wei2024zoomie}, it reserves no logic on the PL fabric, and (ii) PCAP interface reduces integration effort and simplifies deployment on FPGA SoCs. The checkpoint becomes a trusted snapshot of the system state at that verified point.}.~\mytitle~repurposes these primitives of ~\cite{malik2025epoch} from preemption to fault tolerance by (i) introducing hierarchical twins for precise fault detection and localization, (ii) triggering recovery \emph{only} upon detected divergence, and (iii) limiting readback to active PL regions to reduce partial reconfiguration (PR) related read/write traffic and area overhead.

\subsection{Fault Detection and Recovery}
\mytitle's operation involves three phases: (1) normal operation with periodic validation (against \texttt{Rabbit's} output), (2) fault localization (via \texttt{Tortoise}), and (3) state correction with in-place recovery.
\subsubsection{\textbf{Normal Operation with Periodic Validation}} 
PL accelerator executes its operation on the input data. At checkpoint intervals (\textit{e.g.,} after each layer),~\mytitle~captures PL state and offloads it to off-chip memory~\cite{malik2025epoch}. Concurrently,~\mytitle~invokes \texttt{Rabbit} (coarse-grained functional twin) with the same input batch to compute the expected outputs $(O_{\texttt{Rabbit}})$.~\mytitle~then compares $O_{\text{\texttt{Rabbit}}}$ against actual PL outputs $(O_{\text{PL}})$. If $|O_{\text{\texttt{Rabbit}}} - O_{\text{PL}}| < \epsilon$ (within tolerance $\epsilon$), the system continues its normal operation. However, if $|O_{\text{\texttt{Rabbit}}} - O_{\text{PL}}| \geq \epsilon$, a potential fault is flagged, and~\mytitle~transitions to the second phase.
\subsubsection{\textbf{Fault Localization}}
Upon detecting divergence between PL and \texttt{Rabbit}'s output—indicative of a fault—\mytitle~loads the most recent checkpoint (captured at cycle `$t_{capture}$') into \texttt{Tortoise}'s memory, including all registers, control signals, and the input sequence processed by the PL from cycle $t_{capture}$ to the current faulty cycle $(t_{\text{fault}})$. \texttt{Tortoise} then starts emulating it forward, cycle-by-cycle, computing the expected hardware state evolution. At $t_{\text{fault}}$, \texttt{Tortoise}  produces the fault-free values $(O_{\text{\texttt{Tortoise} }})$ that the PL accelerator would have produced in the absence of the fault. After computing the fault-free output,~\mytitle~transitions to the third phase.
\subsubsection{\textbf{State Correction and Recovery}}
\vspace{-0.25em} \mytitle~extracts the corrected state $O_{{\texttt{Tortoise}}}$ from \texttt{Tortoise} to construct a PR bitstream $(t_{restore})$ to write $O_{{\texttt{Tortoise}}}$ state to the corresponding PL registers and memories\footnote{\mytitle~applies this runtime generated bitstream via PCAP, overwriting the corrupted PL state with the corrected values~\cite{malik2025epoch}.}. The PL accelerator resumes execution from the corrected state, continuing its operation \emph{without} full re-execution.

Replay cost may scale with both the checkpoint distance ($\Delta$) and the complexity of the emulated accelerator. Therefore, \texttt{Tortoise} serves as an \textit{infrequent} exception-path mechanism rather than a \textit{continuously} active companion. However, state replay over only $\Delta$ cycles remains less expensive than full re-execution, and the worst-case recovery latency stays bounded by checkpoint spacing and the execution rate of the \texttt{Tortoise}.

\begin{figure}[t]
	\includegraphics[width=0.495\textwidth]{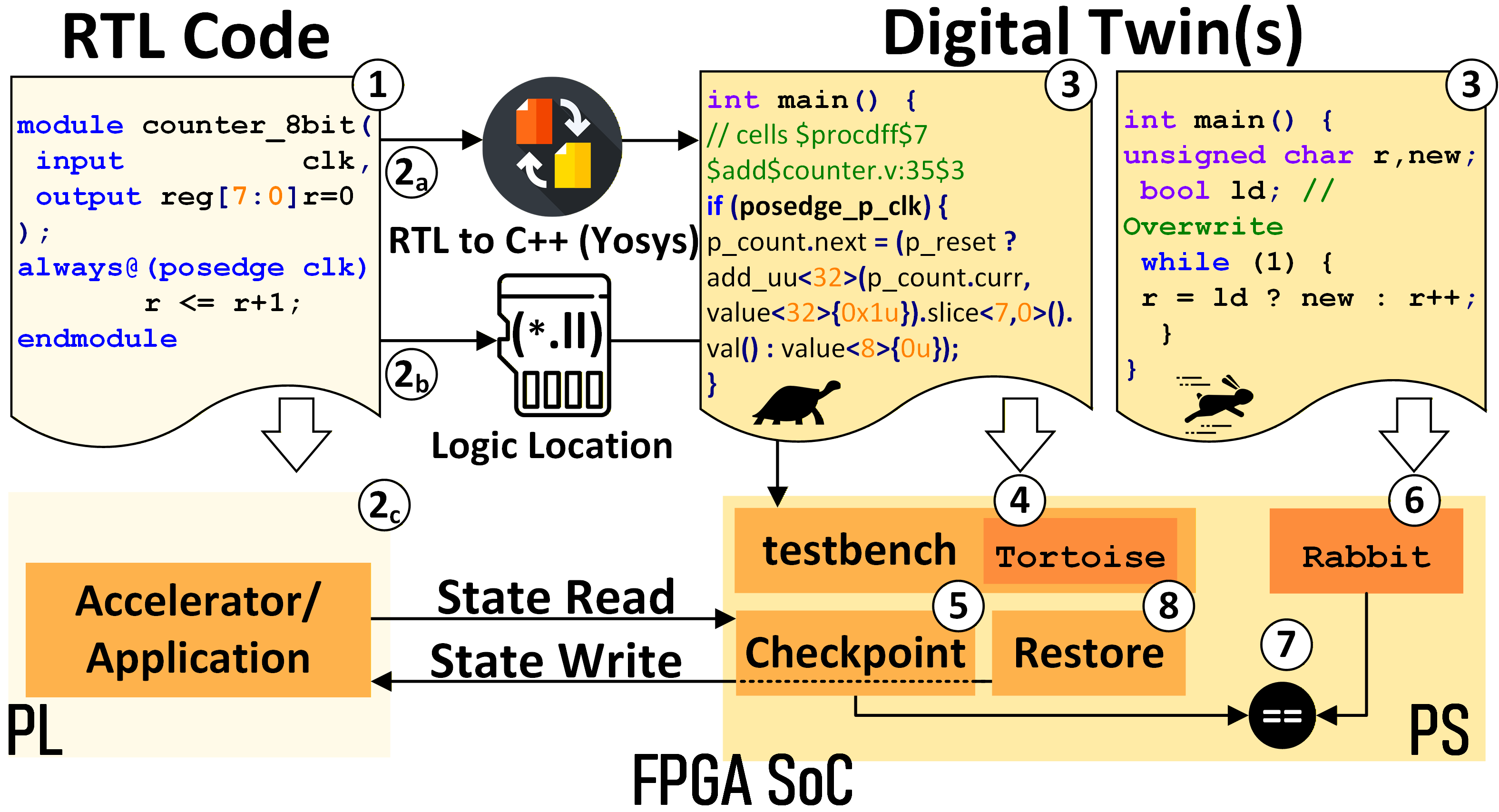}
	\vspace{-1.5em}
	\caption{\mytitle~has eight steps: \ding{172} synthesis of RTL level code to generate bitstream and the logic location (*.ll) file. \ding{173}$_{a}$ conversion of RTL to a gate-level equivalent model (\texttt{Tortoise}) using Yosys~\cite{Yosys,YosysCXXRTL} for fine-grained recovery, \ding{173}$_{b}$ one-time export of the *.ll file, and \ding{173}$_{c}$ programming PL with the generated bitstream. \ding{174} writing and adaptation of \texttt{Rabbit} and \texttt{Tortoise} for PS, respectively. \ding{175} construction of a C++ testbench to instantiate \texttt{Tortoise}. \ding{176} periodic PL state check-pointing. \ding{177} invocation of \texttt{Rabbit} for quick fault detection 
    and \ding{178} result comparison between \texttt{Rabbit} and PL output. \ding{179} Upon detecting a fault, \texttt{Rabbit} triggers \texttt{Tortoise} to emulate fault-free execution, concluding fault mitigation and state recovery.}
\vspace{-1.5em}
	\label{fig: Process}
\end{figure}

\vspace{-0.4em}
\begin{tcolorbox}
\noindent\textbf{Mid-execution self-repair.}
Instead of halting on faults (DMR) or rolling back and re-running everything from scratch (checkpointing),~\mytitle~can \textit{jump-start} from the latest checkpoint and fast-forward to a corrected state, \textit{{e.g.,}} a fault at $t = t_{capture} + \Delta$ requires emulating only $\Delta$ cycles, not a full re-execution.
\end{tcolorbox}
\noindent{\textbf{Coverage Scope and Guarantees.}
\mytitle’s fault coverage is (intentionally) bounded by its checkpointing and validation policy. Under the assumed transient multi-bit fault model and deterministic accelerator execution,~\mytitle~guarantees detection of faults that propagate to monitored outputs within $\Delta$ cycles, where $\Delta$ is defined by the checkpoint spacing\footnote{Faults that do not manifest within the monitored outputs or fall outside the protected computation subset \textit{may} remain undetected.}. This design choice reflects a trade-off between coverage and overhead, prioritizing high-impact fault regions identified through empirical profiling. At present,~\mytitle~does not provide formal end-to-end coverage guarantees. As in prior work, we evaluate coverage through empirical fault modeling and injection, a practical methodology that has been shown to meaningfully capture protection effectiveness against several widely studied attacks. We therefore adopt this established evaluation path for selective accelerator protection, while noting that stronger formal guarantees are still an open research problem.~\cite{natella2016swifi,tyagi2023thales,alvaro2015ldfi}.

%% file: 5.results.tex
\section{Experimental Setup and Results}\label{Sec:Results}
\noindent\textbf{Test-Platform.} We evaluated~\mytitle~on the AMD/Xilinx Zynq-$7000$ (XC$7$Z$020$), which represents a widely deployed platform in safety-critical domains, powering multi-camera advanced driver assistance systems (ADAS) in autonomous vehicles~\cite{xilinx_adas_automated_driving,amd_automotive_zynq_7000_xa}, intelligent traffic management using edge AI~\cite{gong2023edge}, high-availability industrial IoT networks with zero-downtime redundancy~\cite{avnet_zero_downtime_iot}, aerospace~\cite{DMR}, and defense avionics~\cite{xilinx_functional_safety_brief}. On this platform, we evaluated several canonical benchmarks of varying complexity. All accelerators were written in Verilog and synthesized with Vivado $2024.2$. The~\mytitle~framework uses C/C++ on the PS via Vitis SDK, and \texttt{Tortoise} models are generated with Yosys CXXRTL $0.50$~\cite{YosysCXXRTL,Yosys}. 

\noindent\textbf{Experimentation flow.} Fig.~\ref{fig: Process} presents the eight steps that capture the essence of the~\mytitle~framework and the flow of our experiments. While the figure shows snippets for the $8$-bit upcounter example, the steps are universally applicable for any accelerator. \ding{172} The accelerator code is written in RTL language. At this stage, the accelerator is mapped to a specific location using a partition block to track its position. \ding{173$_{a}$} The RTL code is converted into a C++ model using the Yosys CXXRTL backend.
\ding{173$_{b}$} The RTL is synthesized with the logic location file generation setting enabled (later exported to the SD card attached to the PS). This enables~\mytitle~to identify the contents of each logic element and combine their bit values to create complete and consistent state checkpoints. \ding{173$_{c}$} The bitfile is programmed onto the PL, and execution begins.
\input{Tables/Table1-Utilization}
\ding{174} A testbench calls the C++ model generated by Yosys.
\ding{175} The testbench provides the necessary input values and handles the initialization process.
\ding{176} \mytitle~opportunistically captures the state of the accelerator running on the PL and saves it in off-chip DDR memory~\cite{malik2025epoch}.
\ding{177} \texttt{Rabbit} is periodically executed, computing an independent system state. \ding{178} To check if the data values of the PL diverge from the expected values, the system state (computed by the \texttt{Rabbit}) is compared against the output of the PL. \ding{179} In case of a fault, \texttt{Tortoise} runs for a few $\Delta$ cycles, computing the latest fault-free state, which is then restored onto the PL, completing fault mitigation and recovery.

\input{Tables/Table2-Latencies}
\subsection{Fault Injection Setup}\label{FI_Setup}
\input{Tables/Table3-Energy_Area}

To emulate transient faults (without relying on radiation sources or system glitching equipment), we adopted a readback-level fault injection technique, as established in prior works~\cite{zhang2011new,abou2012error,garcia2020autonomous,salvador2023cycle,chen2021hashtag,malik2020isolation}.
During \textit{normal} execution, the PL accelerator processes inputs, and~\mytitle~periodically captures checkpoints via configuration/state readback~\cite{malik2025epoch}. To \emph{accelerate} fault campaigns and ensure rapid, repeatable exposure to fault presence, we inject faults by deliberately corrupting the readback image \emph{after} checkpoint acquisition but \emph{before} the subsequent integrity comparison against \texttt{Rabbit} (Sec.~\ref{Stage_1}). Specifically,~\mytitle~XORs selected bit positions with $1$, thereby emulating transient bit flips, \textit{e.g.,} flipping bit \#$3$ in an $8$-bit register holding $0b1011\textbf{0}101$ produces $0b1011\textbf{1}101$.

Crucially, this injection point is a \emph{methodological convenience} rather than a limitation of the~\mytitle. Corrupting readback data is used only to speed up and systematize fault injection (location/rate sweeps) without requiring invasive hardware fault sources.~\mytitle’s operational principle remains unchanged—when \emph{live} PL faults occur naturally (or are induced by physical means), their effects are captured in its checkpoint state/output.~\mytitle~detects this by comparing the checkpoint-derived state/output against \texttt{Rabbit}. Thus, while our campaign uses readback corruption to emulate the \emph{presence} of a fault efficiently, the same checkpoint-and-compare pipeline is fully capable of detecting faults originating in the live PL computation.




\vspace{-.5em}
\subsection{Benchmarks}
To evaluate~\mytitle, we implemented various canonical designs, inspired from prior works~\cite{malik2026preemption,reagen2014machsuite,malik2025precision}. Table~\ref{tab:Area_Utilization} reports the resource utilization for each of these designs and the digital twins (in terms of code size).
The considered workloads span a broad spectrum of microarchitectural characteristics: (i) a minimal $8$-bit upcounter, used as a sanity check for checkpoint–restore and fault-detection correctness under controlled injections (see Sec.~\ref{FI_Setup}), (ii) cryptographic accelerators (SHA-$256$ and AES-$128$) that exercise~\mytitle~on control-dominated, multi-stage datapaths with finite-state-machine (FSM) control, where faults are injected during active computation to emulate transient cryptographic fault scenarios representative of practical fault-injection attacks~\cite{secworks_sha256,malik2025precision,malik2025CRAFT}, and (iii) representative signal-processing and machine-learning kernels, including dense and Conv2D layers, single and multi-head ($4$) attention layer, general matrix multiplication (GEMM), fast Fourier transform (FFT-64), and digital filters (finite impulse response (FIR)/Conv1D and infinite impulse response (IIR).

For the edge-AI workloads, implementations follow an $8$-bit quantized, systolic-array-style organization (inspired by the Google Edge TPU) to reflect resource-constrained inference settings~\cite{TPU}. Faults are injected into parameter storage and accumulation state (\textit{e.g.,} weights, accumulator registers), consistent with adversarial perturbations of NN parameters in edge deployments~\cite{krautter2018fpgahammer}.

For all benchmarks, \texttt{Rabbit} (Twin \#1) was implemented in C as a behavioral model reflecting the functionality of the corresponding hardware design, while \texttt{Tortoise} (Twin \#2) was generated from each design’s Verilog description using the CXXRTL backend to produce a cycle-accurate gate-level C++ model. Table~\ref{tab:Latencies} presents the execution latencies for the benchmarks on both the PL and the digital twins running on the PS. It also reports the speed ratio relative to the PL, defined as `S' = Twin speed/PL speed.

While \texttt{Rabbit} and \texttt{Tortoise} can be regarded as redundancies in the PS, their overhead is confined to code memory footprint and modest additional PS execution, both of which are easily absorbed by the PS budget. By contrast, DMR in the PL requires a second copy of the accelerator datapath and its on-chip memories, roughly doubling fabric utilization and increasing static and dynamic energy consumption. From Table~\ref{tab:Area_and_Power_Comparison}, this advantage is visible across all case studies: \textbf{DMR\footnote{We use DMR as a baseline because it provides a clear, widely understood reference for always-on hardware replication.  We \textit{intentionally} exclude partial redundancy as a separate baseline because it does not generalize across designs: each new accelerator (or model) requires a new profiling step to identify the protected subset and a corresponding hardware redesign (re-synthesis/re-implementation) to replicate the selected top-$k$ neurons. By contrast, \mytitle\ achieves the same selective protection through the software stack \& retains flexibility across designs without increasing PL area.} increased PL area (LUT/FF) usage by approximately $2.0$–$2.25\times$ and power consumption from $2.98$ W to $5.79$ W ($\approx 89.1$\%), on average}.
\textbf{~\mytitle~incurs a $t_{capture}$ latency of $31.46$~ms, state validation ($\epsilon$) of $39.3$ms, and fault localization $t_{fault}$ of $31.91$ms. By contrast, the actual fault-recovery step $t_{restore}$ that writes the corrected state back to the PL via PCAP completes in only $5.7$ms.} 

The $t_{capture}$ and $t_{fault}$   phases ($\approx$$30$--$40$ ms) are dominated by conservative software implementation choices and PCAP bandwidth, and they are triggered only on rare fault events, leaving steady-state accelerator throughput unaffected. Our evaluation indicates that~\mytitle~can replace continuous hardware redundancy with an on-demand software recovery path while preserving a favorable overhead profile. The resulting downtime from fault detection to PL state recovery remains within only a few milliseconds, which is consistent with typical control-loop deadlines in edge-AI settings~\cite{UG909,perez2019low,kay2015real}. In general, ~\mytitle~is well-suited for resource-constrained FPGA SoCs in which brief recovery stalls are acceptable and eliminating continuous redundancy yields greater overall value.


\subsection{Fault Coverage Across Dense Layer}
\label{sec: Fault Coverage }

\begingroup
\definecolor{CaseStudyColor}{HTML}{000000} 
\color{CaseStudyColor}
Using the percentile-sweep procedure (see Sec.~\ref{Stage_2}), we ranked the dense layer MAC operations by their absolute contribution magnitude $|x\!\cdot\!w|$ and injected multiple bit flips per sampled MAC to measure the fraction of trials resulting in misclassification. The results show that outcome-altering faults are heavily concentrated in the highest-percentile MACs. The smallest top slice capturing all observed misclassifications corresponds to \textbf{30\% of MACs in layer~6, {$\sim$17\%} in layer~7, and {$\sim$18\%} in layer~8}. When aggregated across all dense layers, these findings indicate that auditing approximately \textbf{25\% of dense-layer MACs}—or only \textbf{5\% of all network MACs}—is sufficient to capture all observed misclassifications for the evaluated model under the adopted fault-injection campaign.

Guided by prior work~\cite{b1,b4,b2,b7,b6,gongye2024one,rakin2019bit} and our LeNet-5 case study, we show that restricting comparisons to the \emph{dense/logit} layers and auditing only the highest-magnitude MACs captures nearly all outcome-altering faults. Focusing on the \emph{edge AI} workload, we show that by auditing the top $30\%$ of neurons~\mytitle~provides sufficient fault coverage. Timing profile of dense layer from Table~\ref{tab:Latencies} shows that PL computes the entire layer in approximately $43.54~\mu$s, corresponding to about $10.62$~ns per MAC operation. By contrast, \texttt{Rabbit}'s complete execution on PS processes takes about $585~\mu$s, or $0.14~\mu$s per MAC. Given the PL’s total layer time $t_{\text{PL,layer}}\approx43.54~\mu$s (the time for the PL to compute one dense layer) and the PS per-MAC latency $t_{\text{PS,MAC}}\approx0.14~\mu$s, the PS can recompute at most:

\vspace{-1em} 
\[
C_{\text{MAC}} \;=\; \frac{t_{\text{PL,layer}}}{t_{\text{PS,MAC}}}
\;\approx\; \frac{43.54~\mu\text{s}}{0.14~\mu\text{s}/\text{MAC}}
\;\approx\; 311~\text{MACs}
\]
\vspace{-0.65\baselineskip}
\\\noindent Where $C_{\text{MAC}}$ denotes the number of MACs the PS can verify in-line while the PL completes the layer. This defines the real-time computational budget for in-line comparisons, showing that the PS can verify roughly $7.6$\% of all $4096$ MAC operations per layer. Therefore, selectively checking the most critical $30$\% of neurons (those contributing the largest MAC magnitudes) remains both computationally feasible and sufficient for effective fault coverage. If the PS recomputed the entire dense layer, it would process $4096-311=3785$ MACs that overlap ($MAC_{\text{overlap}}$) with PL execution and the corresponding PS verification time ($t_{\text{PS,full}}$) would be:
\vspace{-0.25em}
\[
t_{\text{PS,full}} \;=\; MAC_{\text{overlap}} \times t_{\text{PS,MAC}}
\;=\; 3785 \times 0.14~\mu\text{s} \;\approx\; 530~\mu\text{s},
\]
\noindent Where $t_{\text{PS,full}}$ is the PS time to verify the non-overlapping MACs. This implies a complete PS-side verification would stall PL operation for roughly $0.53$~ms if isolation is required—undesirable for real-time edge inference with tight latency margins. To avoid such downtime,~\mytitle~employs an importance-aware selection policy that ranks dense-layer MAC operations by their contribution magnitude and verifies only the most influential subset. Let $k$ denote the selected fraction (default $k{=}0.30$ for $30\%$), $N_{\text{sel}}$ the number of selected MACs, and $t_{\text{PS,sel}}$ the PS time to verify them:
\vspace{-0.5em}
\[
\begin{aligned}
N_{\text{sel}} &= k \times 4096 \;=\; 0.3 \times 4096 \;=\; 1228~\text{MACs},\\
t_{\text{PS,sel}} &= N_{\text{sel}} \times t_{\text{PS,MAC}}
\;=\; 1228 \times 0.14~\mu\text{s} \;\approx\; 0.17~\text{ms}.
\end{aligned}
\]
\vspace{-0.20em}
By auditing only the top $30\%$ of MACs, ~\mytitle~reduces the PL isolation window from $530~\mu\text{s}$ to $170~\mu\text{s}$–a nearly threefold improvement–while focusing validation precisely where computation errors are most likely to impact inference accuracy. We set $k{=}30\%$ as the default selection threshold because it (i) aligns with magnitude-aware robustness improvements reported in prior work~\cite{b6}, (ii) matches the selective-protection coverage versus overhead trade-offs demonstrated in larger CNNs~\cite{b2}, and (iii) fits within the PS timing budget, allowing a $\sim 0.17$~ms audit burst (as shown above).

These results should be viewed as workload-specific empirical findings from the evaluated LeNet-5-style dense-layer case study under the assumed multi-bit transient fault model. Accordingly, the $30$\% threshold should not be treated as a universal value for other neural networks or accelerator designs. Instead, it shows that~\mytitle~can leverage measured fault sensitivity to reduce validation cost while maintaining high observed coverage in \textit{any} representative edge-AI setting.

\endgroup

%% file: Tables/Table1-Utilization.tex
\begin{table}[t]
\setlength{\tabcolsep}{3.9pt}
\centering
\small
\caption{Resource utilization for the FPGA SoC accelerators considered in this work.}
\begin{tabular}{|c|c|c|c|c|c|}

\hline
\textbf{Design} & \textbf{LUTs}  & \textbf{FFs}     & \textbf{\texttt{Rabbit} Size}   & \textbf{\texttt{Tortoise} Size} \\ \hline 
Up-counter & $7$ & $8$  &$77.04$KB &$0.90$MB \\ \hline 

SHA-256 & $668$ & $2160$  &$80.21$KB &$1.01$MB \\ \hline 

Dense layer & $809$ & $1086$  &$80.37$KB &$1.01$MB \\ \hline 

Conv2D layer & $12250$ & $21711$  &$3.24$KB &$2.0$MB \\ \hline 

AES-128 & $1796$ & $406$  &$5.62$KB &$2.5$MB \\ \hline

Attention layer & $4990$ & $5721$  &$3.42$KB &$2.3$MB \\ \hline

FFT-64 & $8656$ & $2129$  &$6.16$KB &$0.48$MB \\ \hline

GEMM & $1182$ & $247$  &$4.18$KB &$12.1$KB \\ \hline

FIR/Conv1D & $832$ & $281$  &$2.26$KB &$51.0$KB \\ \hline

IIR & $862$ & $529$  &$2.68$KB &$94.1$KB \\ \hline
Multi-head($4$) attention & $8105$ & $129$  &$5.77$KB &$639$KB \\ \hline
\end{tabular}
\label{tab:Area_Utilization}
\vspace{-1em}
\end{table}



%% file: Tables/Table2-Latencies.tex
\begin{table}[t!]
\setlength{\tabcolsep}{3.9pt}
\centering
\small
\caption{Latency (in $\mu$s) and the speed ratio comparison for the accelerators on FPGA SoC against PL-based implementation.}
\begin{tabular}{|c|c|c|c|c|}

\hline
\textbf{Design} & \textbf{$O_{\texttt{PL}}$}   & \textbf{$O_{\texttt{Rabbit}}$}       
& \textbf{$O_{\texttt{Tortoise}}$}   & \textbf{Speed Ratio (S)}\\ \hline 
Up-counter & $0.01$ & $0.07$  &$4.815$ &$\approx8\times$/ $482\times$                  \\ \hline
SHA-256 & $0.84$ & $18.0$  &$5264$ &$\approx22\times$/ $6266\times$                   \\ \hline
Dense layer & $43.54$ & $585$  &$6695$ &$\approx14\times$/ $154\times$                  \\ \hline

Conv2D layer & $270.40$ & $143$  &$359127$ &$\approx0.5\times$/ $1302\times$                  \\ \hline

AES-128 & $1.6$ & $3.62$  &$9701$ &$\approx2.3\times$/ $6063\times$                  \\ \hline

Attention layer & $28.63$ & $94.8$  &$19906$ &$\approx3.3\times$/ $696\times$                  \\ \hline

FFT-64 & $3.23$ & $5.83$  &$6942$ &$\approx1.8\times$/ $2150\times$  \\ \hline

GEMM & $0.57$ & $1.28$  &$146$ &$\approx2.2\times$/ $256\times$                  \\ \hline

FIR/Conv1D & $0.41$ & $3.00$  &$893$ &$\approx7.3\times$/ $2178\times$  \\ \hline

IIR & $0.39$ & $9.47$  &$3854$ &$\approx25\times$/ $9882\times$  \\ \hline
Multi-head($4$) attention & $0.25$ & $0.27$  &$993$ &$\approx1.08\times$/ $3972\times$  \\ \hline

\end{tabular}
\label{tab:Latencies}
\vspace{-1em}
\end{table}



%% file: Tables/Table3-Energy_Area.tex
\begin{table}[t!]
\setlength{\tabcolsep}{0.001pt}
\centering
\small
\caption{PL area (given as LUTs/FFs) and PL energy (in Watt) overhead of DMR vs.\ ~\mytitle~on FPGA SoC.}

\begin{tabular}{|c|c|c|c|c|}
\hline
\textbf{\rule{0pt}{2.6ex}Design}
  & \textbf{\shortstack{\rule{0pt}{2.6ex}\mytitle~Area\\(LUTs/FFs)}}
  & \textbf{\shortstack{\rule{0pt}{2.6ex}DMR Area\\(LUTs/FFs)}}
  & \textbf{\shortstack{\rule{0pt}{2.6ex}\mytitle~Power\\(in Watt)}}
  & \textbf{\shortstack{\rule{0pt}{2.6ex}DMR Power\\(in Watt)}} \\
\hline
Up-counter & $7/8$ & $15/18$  &$1.683$ &$1.764$ \\ \hline 
SHA-256 & $729/2160$ & $1592/4364$  &$2.90$ &$5.87$ \\ \hline 
Dense layer & $809/1086$ & $1627/2181$  &$2.97$ &$4.90$ \\ \hline 

Conv2D layer & $12250/21711$ & $24500/43430$  &$3.21$ &$6.52$ \\ \hline 
AES-128 & $1796/406$ & $3595/815$  &$2.13$ &$4.40$ \\ \hline 

Attention layer & $4990/5721$ & $9987/11445$  &$4.25$ &$8.71$ \\ \hline 

FFT-64 & $8656/2129$ & $17415/4260$  &$8.13$ &$16.7$ \\ \hline 

GEMM & $1182/247$ & $2370/499$  &$1.83$ &$3.71$ \\ \hline 

FIR/Conv1D & $832/281$ & $1675/563$  &$1.17$ &$2.37$ \\ \hline 

IIR & $862/529$ & $1730/1060$  &$1.54$ &$2.98$ \\ \hline 

Multi-head($4$) attention & $8105/129$ & $16212/260$  &$1.03$ &$2.15$ \\ \hline 
\end{tabular}
\label{tab:Area_and_Power_Comparison}
\vspace{-2em}
\end{table}





%% file: 6.discussions.tex
\section{Discussions} \label{Sec:Discussions}
We now discuss ~\mytitle’s efficacy and outline steps for future work.

\subsection{Limitations and Scope}
\mytitle~depends on three system-level assumptions: (i) The PS can provide sufficient slack to execute \texttt{Rabbit} at checkpoint boundaries without interfering with higher-priority software tasks. (ii) Fault events remain infrequent enough that \texttt{Tortoise’s} slower recovery path does not affect end-to-end performance. (iii) The target application can tolerate brief interruptions during fault detection and state restoration. These conditions hold for the evaluated prototype and, in turn, identify the class of FPGA SoC deployments for which~\mytitle~is most effective.

\mytitle~currently provides empirical fault-detection and recovery coverage for the evaluated transient fault model, workloads, and checkpoint policy. Therefore, the observed coverage depends on the workload’s sensitivity to injected faults, checkpoint placement, and the portion of architectural state that \texttt{Rabbit} validates and \texttt{Tortoise} reconstructs. Extending~\mytitle~to dynamic workloads with time-varying fault sensitivity or adaptive execution graphs remains future work. However, the present evaluation demonstrates that hierarchical twins can deliver practical self-healing with modest overhead on FPGA SoCs.

\mytitle's current implementation uses workload-specific sensitivity profiling (\textit{e.g.,} top $30$\% of neurons as specified in Sec~\ref{subsec: Checkpoint Selection Policy}) to identify the subset of computations that \texttt{Rabbit} re-executes.
Therefore, it presently provisions for neural networks that can leverage top-$k$ checkpoint selection (where $k$ represents the number of layers amenable to checkpoint boundaries). Networks with highly irregular control flow or dynamic execution graphs may require adaptive checkpoint placement algorithms, which is an area of ongoing research. 
\textbf{Sec.~\ref{sec: Fault Coverage } evaluates fault coverage on a dense layer as a representative, challenging edge-AI workload.} Extending the same per-fault coverage analysis to other accelerators (\textit{e.g.}, SHA-\textit{256}, AES-128, GEMM, attention layer, etc.) can be carried out analogously and is omitted due to space constraints. However, it does \textit{not} alter the main conclusion. The benchmarks employed in our evaluation demonstrate that~\mytitle~generalizes across heterogeneous workloads and quantifies its area/power cost, while the dense layer stress-tests its fault detection and recovery.

\subsection{Digital Twins for Fault Detection}
Digital-twin systems are well established in manufacturing and energy domains but remain largely unexplored for hardware fault tolerance on FPGA SoCs~\cite{bofill2023exploring,hosamo2022review,hosamo2023digital,schluse2018experimentable}. Existing twins operate at coarse time scales (seconds to minutes) and assume abundant computing resources, whereas FPGA faults occur at nanosecond scales on resource-constrained edge platforms~\cite{xia2024digital}.~\mytitle~addresses this gap by tailoring hierarchical digital twins to FPGA SoCs, enabling explicit speed–accuracy trade-offs and opportunistic PS execution without contending for PL resources.

The hierarchical twin design plays a central role in scalability because it avoids continuous invocation of the fine-grained twin. However, as system complexity increases, the cost gap between \texttt{Rabbit’s} coarse-grained validation and \texttt{Tortoise’s} replay-based recovery \textit{may} grow. Under these conditions, selective activation, partial-state replay, and modular decomposition become increasingly important for preserving responsiveness without sacrificing recovery capability. Evaluating these tradeoffs on larger SoC platforms, therefore, remains a key direction for future work.

\subsection{Conditions for Correct Recovery}

In case of faults, correct restoration depends on deterministic replay from the most recent trusted checkpoint to the point of fault detection.~\mytitle~therefore assumes deterministic replay over bounded execution windows.~\mytitle~does \textit{not} claim correctness in the presence of untracked nondeterminism, such as asynchronous software intervention, uncontrolled external input streams, or timing-dependent side effects outside the state replay/restoration window.

\subsection{Safe Update and State Injection}

\mytitle~applies recovery updates only after the design enters a controlled quiescent state. 
~\mytitle~halts system clock(s) only at safe synchronization points using a handshake mechanism\footnote{A transmit-receive acknowledgment handshake between the PS and PL, driven by the slowest system clock, enables design suspension at safe synchronization points.}, thereby preserving the tick relationship among clock domains and preventing a `torn' state.~\mytitle's mechanism also promotes completing in-flight transactions before recovery begins, thereby avoiding protocol violations and inconsistent states during partial reconfiguration. 

To avoid collateral modification,~\mytitle~\textit{only} changes the configuration bits associated with the target state elements identified by the offline state-to-frame mapping (see Sec ~\ref{Fault Recovery and Restore-Path}). This implementation-specific mapping localizes each update to the relevant frame addresses and bit positions of the target state elements. Consequently,\-~\mytitle~does \textit{not} modify unrelated LUT initialization bits or routing resources. During fault recovery, the injected update remains confined to the intended architectural state.

\subsection{Fault Recovery and Restore-Path}
\label{Fault Recovery and Restore-Path}

\mytitle~builds on the checkpoint-and-resume mechanism of~\cite{malik2025epoch} and repurposes it for fault recovery. Presently,~\mytitle~restores architectural state that satisfies three conditions: it is (i) observable in the deployed design, (ii) reproducible by \texttt{Tortoise} from the same RTL description used to generate the final bitstream, and (iii) writable through the implementation-specific restore path. This state includes flip-flops, pipeline registers, control registers, and other sequential elements as represented in \texttt{Tortoise's} cycle-accurate model. Recovery, therefore, targets a state that is checkpoint-visible, replay-reproducible, and restorable through configuration updates.


The mapping from \texttt{Tortoise}-visible state to PL's configuration frames—and vice versa—is derived offline for each finalized implementation from the logic location (\texttt{*.ll}) file. The \texttt{*.ll} file is produced during bitstream generation and contains the physical locations of context-relevant state-bearing elements, such as flip-flops, thereby identifying the configuration frames that hold their implemented state, enabling precise state extraction and restoration~\cite{malik2025epoch}.
Because the mapping is derived from the finalized placed-and-routed design, it is bitstream-specific rather than synthesis-independent, as outlined in prior works~\cite{attia2020statemover,malik2025epoch,xilinx2023readback}. It therefore remains valid only for a fixed implementation and must be regenerated whenever RTL logic, synthesis, placement and routing, or toolchain versions modify the implemented layout.

\subsection{Scalability}
\mytitle’s scalability is governed by the interaction between checkpoint spacing, accelerator complexity, and available PS resources. The cost of \texttt{Rabbit} scales with the fraction of computation selected for validation, while \texttt{Tortoise}’s recovery latency scales with checkpoint spacing and the structural complexity of the accelerator. As accelerator size increases, larger state and deeper pipelines \textit{may} increase replay cost. However, this overhead remains bounded by checkpoint intervals rather than full re-execution.~\mytitle~is therefore best suited for providing fault tolerance in designs where chip area is limited—making full hardware copies, such as DMR/TMR, infeasible—and where fault events are infrequent, allowing for selective validation to capture high-impact errors \textit{without} proportional increases in area and power overhead.

%% file: 7.conclusion.tex
\section{Conclusion}
This work introduced~\mytitle, a hierarchical digital-twin framework for self-healing FPGA SoCs in edge AI applications. By decoupling fast detection (coarse-grained \texttt{Rabbit}) from precise recovery (fine-grained \texttt{Tortoise}),~\mytitle~provides fault tolerance without the PL area and power overheads of DMR/TMR. Opportunistic twin invocation and checkpoint-driven recovery enable energy-efficient operation on resource-constrained devices. Experiments on an AMD/Xilinx Zynq-$7000$ across multiple benchmarks demonstrate successful detection and correction of transient faults. 
Future work includes scaling~\mytitle~to larger accelerators, studying adaptive checkpoint placement under dynamic workloads, and developing more selective replay and recovery mechanisms to manage checkpoint size, twin latency, and PS utilization.

\vspace{1em}
\noindent  